\RequirePackage{fix-cm}
\documentclass[smallextended]{svjour3}       
\smartqed  
\usepackage{amsmath}
\usepackage{amsfonts}
\usepackage{latexsym}
\usepackage{graphicx}
\usepackage{color}
\usepackage{mathptmx}      

\usepackage{makecell}
\usepackage[breaklinks]{hyperref}
\usepackage{url}
\usepackage{breakurl}
\usepackage{todo}
\makeatletter
\g@addto@macro{\UrlBreaks}{\UrlOrds}
\makeatother
\begin{document}

\title{Is there a role for statistics in artificial intelligence?
}


\author{Sarah Friedrich$^{1}$ \and Gerd Antes$^2$ \and Sigrid Behr$^3$ \and Harald Binder$^4$ \and Werner Brannath$^5$ \and Florian Dumpert$^6$  \and Katja Ickstadt$^7$ \and Hans Kestler$^{8}$ \and  Johannes Lederer$^9$ \and Heinz Leitg\"ob$^{10}$ \and Markus Pauly$^7$ \and Ansgar Steland$^{11}$ \and Adalbert Wilhelm$^{12}$ \and Tim Friede$^1$\footnote{Corresponding author:  \email{tim.friede@med.uni-goettingen.de}}
}

\authorrunning{Friedrich et al.} 

\institute{$^1$ Department of Medical Statistics, University Medical Center G\"ottingen, Germany
	\and
	$^2$ Cochrane Germany, Universit\"atsklinikum Freiburg, Germany
           \and
          $^3$ Novartis Pharma AG, Basel, Switzerland
          \and
          $^4$ Faculty of Medicine and Medical Center, University of Freiburg, Germany
          \and
          $^5$ Competence Center for Clinical Trials and Institute of Statistics, University Bremen, Germany
          \and
          $^6$ Federal Statistical Office of Germany
          \and
          $^7$ Department of Statistics, TU Dortmund University, Germany
          \and
          $^{8}$ Institute of Neural Information Processing, Ulm University, Germany	
          \and
          $^9$ Faculty of Mathematics, Ruhr-Universit\"at Bochum, Germany
          \and
          $^{10}$ Universit\"at Eichst\"att-Ingolstadt, Germany
          \and
          $^{11}$ RWTH Aachen, Germany
          \and
          $^{12}$ Jacobs University Bremen, Germany
}

\date{Received: date / Accepted: date}

\maketitle

\begin{abstract}
The research on and application of artificial intelligence (AI) has triggered a comprehensive scientific, economic, social and political discussion. Here we argue that statistics, as an interdisciplinary scientific field,  plays a substantial role both for the theoretical and practical understanding of AI and for its future development. Statistics might even be considered a core element of AI. With its specialist knowledge of data evaluation, starting with the precise formulation of the research question and passing through a study design stage on to analysis and interpretation of the results, statistics is a natural partner for other disciplines in teaching, research and practice.
This paper aims at contributing to the current discussion by highlighting the relevance of statistical methodology in the context of AI development. In particular, we discuss contributions of statistics to the field of artificial intelligence concerning methodological development, planning and design of studies, assessment of data quality and data collection, differentiation of causality and associations and assessment of uncertainty in results. Moreover, the paper also deals with the equally necessary and meaningful extension of curricula in schools and universities.

\keywords{Statistics \and Artificial Intelligence \and Machine Learning \and Data Science}
 \subclass{68T01 \and 62-02}
\end{abstract}

\section{Introduction}
\label{intro}

The research on and application of artificial intelligence (AI) has triggered a comprehensive scientific, economic, social and political discussion. Here we argue that statistics, as an interdisciplinary scientific field, plays a substantial role, both for the theoretical and practical understanding of AI and for its further development. 

Contrary to the public perception, AI is not a new phenomenon. AI was already mentioned in 1956 at the Dartmouth Conference \cite{[54],[53]}, and the first data-driven algorithms such as Perceptron \cite{[77]} and backpropagation \cite{[78]} were developed in the 50s and 60s. The Lighthill Report in 1973 made a predominantly negative judgment on AI research in Great Britain and led to the fact that the financial support for AI research was almost completely stopped (the so-called \emph{first AI winter}). The following phase of predominantly knowledge-based development ended in 1987 with the so-called \emph{second AI winter}. In 1988, Judea Pearl published his book `Probabilistic Reasoning in Intelligent Systems', for which he received the Turing Award in 2011 \cite{[80]}. From the beginning of the 1990s, AI has been developing again with major breakthroughs like Support Vector Machines \cite{[81]}, Random Forest \cite{[82]}, Bayesian Methods \cite{[136]}, Boosting and Bagging \cite{[97],[98]}, Deep Learning \cite{[83]} and Extreme Learning Machines \cite{[99]}. 

Today, AI plays an increasingly important role in many areas of life. International organizations and national governments have currently positioned themselves or introduced new regulatory frameworks for AI. Examples are, among others, the AI strategy of the German government \cite{[1]} and the statement of the Data Ethics Commission \cite{[107]} from 2019.
Similarly, the European Commission recently published a white paper on AI \cite{europ}.
 Furthermore, regulatory authorities such as the US Food and Drug Administration (FDA) are now also dealing with AI topics and their evaluation. In 2018, for example, the electrocardiogram function of the Apple Watch was the first AI application to be approved by the FDA \cite{[84]}.

There is no unique and comprehensive definition of \emph{artificial intelligence}. Two concepts are commonly used distinguishing \emph{weak} and \emph{strong AI}. Searle (1980) defined them as follows: \emph{`According to weak AI, the principal value of the computer in the study of the mind is that it gives us a very powerful tool. [...] But according to strong AI, the computer is not merely a tool in the study of the mind; rather, the appropriately programmed computer really is a mind [...]'} \cite{Searle1980}. In the following, we adopt the understanding of the German Government, which is the basis of Germany's AI strategy \cite{[1]}. In this sense, (weak) AI \emph{`is focused on solving specific application problems based on methods from mathematics and computer science, whereby the developed systems are capable of self-optimization'} \cite{[1]}.
An important aspect thus is that AI systems are self-learning. Consequently, we will focus on the data-driven aspects of AI in this paper. In addition, there are many areas in AI that deal with the processing of and drawing inference from symbolic data \cite{[55]}, which will not be discussed here.

As for AI, there is neither a single definition nor a uniform assignment of methods to the field of machine learning (ML) in literature and practice. Based on Simon's definition from 1983 \cite{[2]}, learning describes changes of a system in such a way that a similar task can be performed more effectively or efficiently the next time it is performed. 

Often the terms AI and ML are mentioned along with \emph{Big Data} \cite{gudivada2015} or \emph{Data Science}, sometimes even used interchangeably. However, neither are AI methods necessary to solve Big Data problems, nor are methods from AI only applicable to Big Data. Data Science, on the other hand, is usually considered as an intersection of computer science, statistics and the respective  scientific discipline. Therefore, it is not bound to certain methods or certain data conditions.

This paper aims at contributing to the current discussion by highlighting the relevance of statistical methodology in the context of AI development and application. 
Statistics can make important contributions to a more successful and secure use of AI systems, for example with regard to
\begin{enumerate}
	\item Design: bias reduction; validation; representativity
	\item Assessment of data quality and data collection: standards for the quality of diagnostic tests and audits; dealing with missing values
	\item Differentiation between causality and associations: consideration of covariate effects; answering causal questions; simulation of interventions
	\item Assessment of certainty or uncertainty in results: Increasing interpretability; mathematical validity proofs or theoretical properties in certain AI contexts; providing stochastic simulation designs; accurate analysis of the quality criteria of algorithms in the AI context
\end{enumerate}

The remainder of the paper is organized as follows: First, we present an overview of AI applications and methods in Section \ref{sec:overview}. We continue by expanding on the points 1.--4. in Sections \ref{design} -- \ref{uncertainty}. Furthermore, we discuss the increased need for teaching and further education targeting the increase of AI-related literacy at all educational levels in Section \ref{edu}. We conclude with Section \ref{summary}.

\section{Applications and Methods of AI}\label{sec:overview}
Depending on the specific AI task (e.~g., prediction, explanation, classification), different approaches are used, ranging from regression to deep learning algorithms. Important categories of AI approaches are supervised learning, unsupervised learning and reinforcement learning \cite{[56]}. Many AI systems learn from training data with predefined solutions such as true class labels or responses. This is called supervised learning, whereas unsupervised learning does not provide solutions. In contrast, reinforcement learning has no predefined data and learns from errors within an agent-environment system, where Markov decision processes from probability theory play an important role. The input data can be measured values, stock market prices, audio signals, climate data or texts, but may also describe very complex relationships, such as chess games. Table 1 gives an overview of examples of statistical methods and models used in AI systems. The models could be further classified as data-based and theory-based models. However, we have omitted this classification in the table.

Even though many of the contributions to AI systems originate from computer science, statistics has played an important role throughout. Early examples occurred in the context of realizing the relationship between backpropagation and nonlinear least squares methods, see, e.g., \cite{warner1996understanding}. Machine learning (ML) plays a distinctive role within AI. For instance, important ML methods such such as random forests \cite{[82]} or support vector machines \cite{[81]}, were developed by statisticians. Others, like radial basis function networks \cite{[106]}, can also be considered and studied as nonlinear regression models. Recent developments such as extreme learning machines or broad learning systems \cite{[104]} have close links to multiple multivariate and ridge regression, i.e. to statistical methods. The theoretical validity of machine learning methods, e.g., through consistency statements and generalization bounds \cite{[105],[57]}, also requires substantial knowledge of mathematical statistics and probability theory. 

\begin{figure}[ht]
	\includegraphics[width=\textwidth]{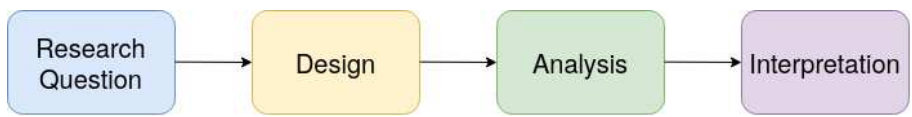}
	\caption{Flow chart of study planning, design, analysis and interpretation}
	\label{fig:flowchart}       
\end{figure}

To capture the role and relevance of statistics, we consider the entire process of establishing an AI application. As illustrated in Figure~\ref{fig:flowchart}, various steps are necessary to examine a research question empirically. For  more details on these steps  see, e.g., \cite{weihs}. Starting with the precise formulation of the research question the process then runs through a study design stage (including sample size planning and bias control) to the analysis. Finally, the results must be interpreted. AI often focuses on the step of data analysis while the other stages receive less attention or are even ignored. This may result in critical issues and possibly misleading interpretations, such as sampling bias or the application of inappropriate analysis tools requiring assumptions not met by the chosen design.

In many applications, the input data for the AI techniques are very high-dimensional, i.e. numerous variables (also called features) are observed with diverse ranges of possible values. In addition, non-linear relationships with complex interactions are often considered for prediction. However, even with sample sizes in the order of millions, the problem of the \emph{curse of dimensionality} arises \cite{[4]}, because data is thin and sparse in a high-dimensional space. Therefore, learning the structure of the high-dimensional space from these thin data typically requires an enormous amount of training data. 

\begin{table}[ht]
	\centering
	\caption{Overview of statistical methods and models used in AI systems \cite{[89],[86],[87],[88]}}
	\label{tab:1}       
	\begin{tabular}{cccc}
		\hline\noalign{\smallskip}
		\textbf{Purpose} & \textbf{ML approach} & \makecell{\textbf{Statistical methods} \& \\ \textbf{models (examples) }} & \textbf{Examples} \\
		\noalign{\smallskip}\hline\noalign{\smallskip}
		Recognizing similarities in data & unsupervised learning & \makecell{Cluster analysis, \\ factor analysis} & \makecell{Personalized medicine \cite{[65]}, \\ customer analysis,\\ development of psychometric tests \cite{[76]} }\\ 	\noalign{\smallskip}\hline
		Prediction of events/conditions & \makecell{Supervised Learning: \\ Regression ML systems} & \makecell{Data-driven \\ model selection} & \makecell{Sales forecast, \\ economic development \cite{[68]},\\
		Weather/climate forecast \cite{[66]},\\  forecast of migration movements \cite{[67]}}\\
		\noalign{\smallskip}\hline
		
		Explanation of events/conditions & \makecell{Supervised Learning: \\ Regression ML systems, \\ interpretable} & \makecell{Theory-based \\ model selection} & \makecell{Risk factors in epidemiology \cite{[69],[70]},\\
			test theory}\\
		\noalign{\smallskip}\hline
		
		Detection of specific objects & \makecell{Supervised Learning: \\ Classification ML systems} & Classification & \makecell{speech and facial recognition \cite{[71],[72]}, \\
			diagnosis and differential diagnosis\\ \cite{[74],[73],[75]}}\\
		\noalign{\smallskip}\hline
		
	\end{tabular}
\end{table}

AI has made remarkable progress in various fields of application. These include automated face recognition, automated speech recognition and translation \cite{[100]}, object tracking in film material, autonomous driving, and the field of strategy games such as chess or go, where computer programs now beat the best human players \cite{[95],[96]}. 

Especially for tasks in speech recognition as well as text analysis and translation, Hidden Markov models from statistics are used and further developed with great success \cite{[94],[101]} because they are capable of representing grammars. Nowadays, automatic language translation systems can even translate languages such as Chinese into languages of the European language family in real time and are used, for example, by the EU \cite{[102]}. Another growing and promising area for AI applications is medicine. Here, AI is used, e.g., to improve the early detection of diseases, for more accurate diagnoses, or to predict acute events \cite{[23],[24]}. Official statistics uses AI methods for classification as well as for recognition, estimation and/or imputation of relevant characteristic values of statistical units \cite{stekhoven2012,shah2014,ramosaj2019,ramosaj2020}. In economics and econometrics, AI methods are also applied and further developed, for example, to draw conclusions about macroeconomic developments from large amounts of data on individual consumer behavior \cite{[129],[130]}.

Despite these positive developments that also dominate the public debate, some caution is advisable. There are a number of reports about the limits of AI, e.g., in the case of a fatal accident involving an autonomously driving vehicle \cite{[6]}. Due to the potentially serious consequences of false positive or false negative decisions in AI applications, careful consideration of these systems is required \cite{[85]}. This is especially true in applications such as video surveillance of public spaces. For instance, a pilot conducted by the German Federal Police at the S\"udkreuz suburban railway station in Berlin has shown that automated facial recognition systems for identification of violent offenders currently have false acceptance rates of 0.67\% (test phase 1) and 0.34\% (test phase 2) on average \cite{[103]}. This means that almost one in 150 (or one in 294) passers-by is falsely classified as a violent offender. In medicine, wrong decisions can also have drastic and negative effects, such as an unnecessary surgery and chemotherapy in the case of wrong cancer diagnoses. Corresponding test procedures for medicine are currently being developed by regulators such as the US FDA \cite{[7]}. 

Finally, ethical questions arise regarding the application of AI systems \cite{[107]}. Apart from fundamental considerations (which decisions should machines make for us and which not?), even a principally socially accepted application with justifiable false decision rates can raise serious ethical questions. This is particularly true if the procedure in use discriminates against certain groups or minorities (e.g., the application of AI-based methods of predictive policing may induce racial profiling \cite{Brantingham2018}) or if there is no sufficient causality.

\section{Study design}
\label{design}

The design of a study is the basis for the validity of the conclusions. However, AI applications often use data that were collected for a different purpose (so-called secondary data). A typical case is the scientific use of routine data. For example, the AI models in a recent study about predicting medical events (such as hospitalization) are based on medical billing data \cite{[60]}. Another typical case are convenience samples, that is, samples that are not randomly drawn but instead depend on `availability'. Well-known examples are online questionnaires, which only reach those people who visit the corresponding homepage and take the time to answer the questions. Statistics distinguishes between two types of validity \cite{[8]}:
\begin{enumerate}
	\item Internal validity is the ability to attribute a change in an outcome of a study to the investigated causes. In clinical research, e.g., this type of validity is ensured through randomization in controlled trials. Internal validity in the context of AI and ML can also refer to avoiding systematic bias (such as systematically underestimated risks).
	\item External validity is the ability to transfer the observed effects and relationships to larger or different populations, environments, situations, etc. In the social sciences (e.g. in the context of opinion research), an attempt to achieve this type of validity is survey sampling, which comprises sampling methods that aim at representative samples in the sense of \cite{[63],kruskal1,kruskal2,kruskal3,kruskal4}. 
\end{enumerate}
 
The naive expectation that sufficiently large data automatically leads to representativity is false \cite{[9],[10]}. A prominent example is Google Flu \cite{[11]}, where flu outbreaks were predicted on the basis of search queries: it turned out that the actual prevalence was overestimated considerably. Another example is Microsoft's chatbot Tay \cite{[12],[13]}, which was designed to mimic the speech pattern of a 19-year-old American girl and to learn from interactions with human users on Twitter: after only a short time, the bot posted offensive and insulting tweets, forcing Microsoft to shut down the service just 16 hours after it started. And yet another example is the recently published Apple Heart Study \cite{[14]}, which examined the ability of Apple Watch to detect atrial fibrillation: there were more than 400,000 participants, but the average age was 41 years, which is particularly problematic in view of atrial fibrillation occurring almost exclusively in people over 65 years of age.

\begin{figure}[ht]
	  \includegraphics[width=\textwidth]{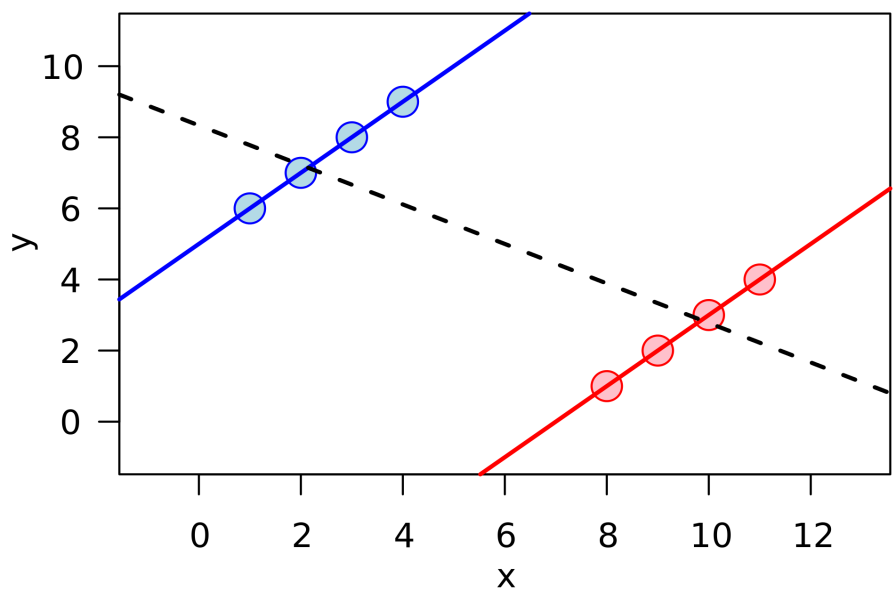}
	\caption{Simpson's paradox for continuous data: a positive trend is visible for both groups individually (red and blue), but a negative trend (dashed line) appears when the data are pooled across groups \cite{[124]}}
	\label{fig:simpson}       
\end{figure}

If data collection is not accounted for, spurious correlations and bias (in its many forms, such as selection, attribution, performance, and detection bias) can falsify the conclusions. A classic example of such falsification is Simpson's paradox \cite{[15]}, which describes a reversal of trends when subgroups are disregarded, see Figure~\ref{fig:simpson}. Further examples are biases inflicted by how the data are collected, such as length time bias.

Statistics provides methods and principles for minimizing bias. Examples include the assessment of the risk of bias in medicine \cite{[16]}, stratification, marginal analyses, consideration of interactions, and meta-analyses, and techniques specifically designed for data collection such as (partial) randomization, (partial) blinding, and methods of so-called optimal designs \cite{karlin1966optimal}. Statistics also provides designs that allow for the verification of internal and external validity \cite{[125],[126],[127]}.

Another important factor for the significance and reliability of a study is the sample size \cite{[9]}. For high-dimensional models, an additional factor is `sparsity' (see the introduction). Through statistical models and corresponding mathematical approximations or numerical simulations, statisticians can assess the potentials and limits of an AI application for a given number of cases or estimate the necessary number of cases in the planning stage of a study. This is not routine work; instead, it requires advanced statistical training, competence, and experience.

Thus, statistics can help in collecting and processing data for subsequent use in AI pipelines. Basic statistical techniques that are relevant for this also include the modeling of data generating processes, restrictions on data sets, and factorial design of experiments, which is a controlled variation of factors that highlights their influence \cite{[17]}. In addition, the various phases in the development of a diagnostic tests are specified in statistics [18]. In many AI applications, however, the final evaluation phase on external data is never reached, since the initial algorithms have been replaced in the meanwhile. Also, statistical measures of quality such as sensitivity, specificity, and ROC curves are used in the evaluation of AI methods. And finally, statistics can help in the assessment of uncertainty (Section \ref{uncertainty}).

\section{Assessment of data quality and data collection} \label{quality}

`Data is the new oil of the global economy.' According to, e.g., the New York Times \cite{[141]} or the Economist \cite{[140]}, this credo echoes incessantly through start-up conferences and founder forums. However, this metaphor is both popular and false. First of all, data in this context corresponds to crude oil, which needs further refining before it can be used. In addition, the resource crude oil is limited. `For a start, while oil is a finite resource, data is effectively infinitely durable and reusable' (Bernard Marr in Forbes, \cite{[142]}). All the more important is a responsible approach to data processing.

\begin{figure}[ht]
	\includegraphics[width=\textwidth]{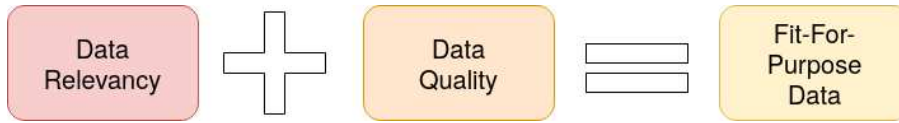}
	\caption{Data relevancy and quality are equivalent components of a fit-for-purpose real-world data set. Figure according to Duke-Margolis \cite{[61]}}
	\label{fig:fit-for-purpose}       
\end{figure}

Ensuring data quality is of great importance in all analyses, according to the popular slogan `Garbage in, garbage out.' As already mentioned in the previous section, we mainly use secondary data in the context of AI. The collection and compilation of secondary data is in general not based on a specific question. Instead, it is collected for other purposes such as accounting or storage purposes. The concept of knowledge discovery in databases \cite{[121]} very clearly reflects the assumption that data are regarded as a given basis from which information and knowledge can be extracted by AI procedures. This notion is contrary to the traditional empirical research process, in which empirically testable research questions are derived from theoretical questions by conceptualizing and operationalizing. The resulting measurement variables are then collected for this specific purpose. In the context of AI, the process of operationalization is replaced by the ETL process: `Extract, Transform, Load' \cite{[122]}. Relevant measurements are to be extracted from the data lake(s), then transformed and finally loaded into the (automated) analysis procedures. The AI procedures are thereby expected to be able to distill relevant influencing variables from high-dimensional data. 

The success of this procedure fundamentally depends on the quality of the data. In line with Karr et al. (2006), data quality is defined here as the ability of data to be used quickly, economically and effectively for decision-making and evaluation \cite{[19]}. In this sense, data quality is a multi-dimensional concept that goes far beyond measurement accuracy and includes aspects such as relevance, completeness, availability, timeliness, meta-information, documentation and, above all, context-dependent expertise \cite{[61],[62]}. In official statistics, relevance, accuracy and reliability, timeliness and punctuality, coherence and comparability, accessibility and clarity are defined as dimensions of data quality \cite{[128]}.

Increasing automation of data collection, e.g., through sensor technology, may increase measurement accuracy in a cost-effective and simple way. Whether this will achieve the expected improvement in data quality remains to be seen in each individual application. Missing values are a common problem of data analyses. In statistics, a variety of methods have been developed to deal with these, including imputation procedures, or methods of data enhancement \cite{rubin1976,seaman2013,vanbuuren2018}. The AI approach of ubiquitous data collection allows the existence of redundant data, which can be used with appropriate context knowledge to complete incomplete data sets. However, this requires a corresponding integration of context knowledge into the data extraction process.

The data-hungry decision-making processes of AI and statistics are subject to a high risk with regard to relevance and timeliness, since they are implicitly based on the assumption that the patterns hidden in the data should perpetuate themselves in the future. In many applications, this leads to an undesirable entrenchment of existing stereotypes and resulting disadvantages, e.g., in the automatic granting of credit or the automatic selection of applicants. A specific example is given by the gender bias in Amazon's AI recruiting tool \cite{[20]}. 

In the triad `experiment - observational study - convenience sample (data lake)', the field of AI, with regard to its data basis, is moving further and further away from the classical ideal of controlled experimental data collection to an exploration of given data based on pure associations. However, only controlled experimental designs guarantee an investigation of causal questions. This topic will be discussed in more detail in Section \ref{causal}.

Exploratory data analysis provides a broad spectrum of tools to visualize the empirical distributions of the data and to derive corresponding key figures. This can be used in preprocessing to detect anomalies or to define ranges of typical values in order to correct input or measurement errors and to determine standard values. In combination with standardization in data storage, data errors in the measurement process can be detected and corrected at an early stage. This way, statistics helps to assess data quality with regard to systematic, standardized and complete recording. Survey methodology primarily focuses on data quality. The insights gained in statistical survey research to ensure data quality with regard to internal and external validity provide a profound foundation for corresponding developments in the context of AI. Furthermore, various procedures for imputing missing data are known in statistics, which can be used to complete the data depending on the existing context and expertise \cite{rubin1976,seaman2013,vanbuuren2018}. Statisticians have dealt intensively with the treatment of missing values under different development processes (non-response, missing not at random, missing at random, missing completely at random \cite{rubin1976,[108]}), selection bias and measurement error.

Another point worth mentioning is parameter tuning, i.e. the determination of so-called hyperparameters, which control the learning behavior of ML algorithms: comprehensive parameter tuning of methods in the AI context often requires very large amounts of data. For smaller data volumes it is almost impossible to use such procedures. However, certain model-based (statistical) methods can still be used in this case \cite{richter2019}.

\section{Distinction between causality and associations}\label{causal}

Only a few decades ago, the greatest challenge of AI was to enable machines to associate a possible cause with a set of observable conditions. The rapid development of AI in recent years (both in terms of the theory and methodology of statistical learning processes and the computing power of computers) has led to a multitude of algorithms and methods that have now mastered this task. One example are deep learning methods, which are used in robotics \cite{[21]} and autonomous driving \cite{[22]}, as well as in computer-aided detection and diagnostic systems (e.g., for breast cancer diagnosis \cite{[23]}), drug discovery in pharmaceutical research \cite{[24]} and agriculture \cite{[25]}. 
With their often high predictive power, AI methods can uncover structures and relationships in large volumes of data based on associations. Due to the excellent performance of AI methods in large data sets, they are also frequently used in medicine to analyze register and observational data that have not been collected within the strict framework of a randomized study design (Section~\ref{design}). However, the discovery of correlations and associations (especially in this context) is not equivalent to establishing causal claims.

An important step in the further development of AI is therefore to replace associational argumentation with causal argumentation. Pearl \cite{[27]} describes the difference as follows: \emph{`An associational concept is any relationship that can be defined in terms of a joint distribution of observed variables, and a causal concept is any relationship that cannot be defined from the distribution alone.'}

\begin{figure}[ht]
	\includegraphics[width=\textwidth]{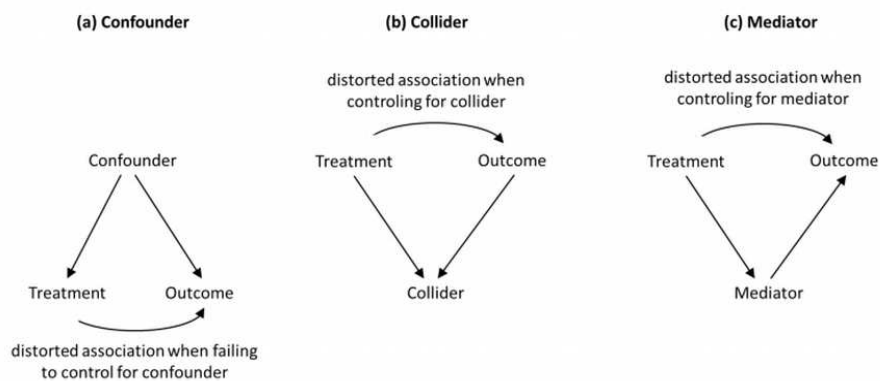}
	\caption{Covariate effects in observational data, according to \cite{[123]}}
	\label{fig:confounder}       
\end{figure}

Even the formal definition of a causal effect is not trivial. The fields of statistics and clinical epidemiology, for example, use the Bradford Hill criteria \cite{[28]} and the counterfactual framework introduced by Rubin \cite{[29]}.
The central problem in observational data are covariate effects, which, in contrast to the randomized controlled trial, are not excluded by design and whose (non-)consideration leads to distorted estimates of causal effects. In this context, a distinction must be made between \emph{confounders}, \emph{colliders}, and \emph{mediators} \cite{[30]}. Confounders are unobserved or unconsidered variables that influence both the exposure and the outcome, see Figure~\ref{fig:confounder}~(a). This can distort the effects of exposure if naively correlated. Fisher identified this problem in his book `The Design of Experiments' published in 1935. A formal definition was developed in the field of epidemiology in the 1980s \cite{[31]}. Later, graphical criteria such as the Back-Door Criterion \cite{[32],[33]} were developed to define the term \emph{confounding}. 

In statistics, the problem of confounding is taken into account either in the design (e.g., randomized study, stratification, etc.) or evaluation (propensity score methods \cite{[34]}, marginal structural models \cite{[35]}, graphical models \cite{[36]}). In this context, it is interesting to note that randomized studies (which have a long tradition in the medical field) have recently been increasingly used in econometric studies \cite{[38],[37],[119]}. In the case of observational data, econometrics has made many methodological contributions to the identification of treatment effects, e.g., via the potential outcome approach \cite{[132],[133],[134],[29],[131]} as well as the work on policy evaluation \cite{[135]}.

In contrast to confounders, colliders and mediators lead to distorted estimates of causal effects precisely when they are taken into account during estimation. Whereas colliders represent common consequences of treatment and outcome (Figure~\ref{fig:confounder}~(b)), mediators are variables that represent part of the causal mechanism by which the treatment affects the outcome (Figure~\ref{fig:confounder}~(c)). Especially in the case of longitudinal data, it is therefore necessary to differentiate in a theoretically informed manner which relationship covariates in the observed data have with the treatment and outcome, thus avoiding bias in the causal effect estimates by (not) having taken them into account.

By integrating appropriate statistical theories and methods into AI, it will be possible to answer causal questions and simulate interventions. In medicine, e.g., questions such as `What would be the effect of a general smoking ban on the German health care system' can then be investigated and reliable statements can be made, even without randomized studies. Of course, this would not be possible here. Pearl's idea goes beyond the use of ML methods in causal analyses (which are used, for example, in connection with targeted learning \cite{[39]} or causal random forest \cite{[40]}). His vision is rather to integrate the causal framework \cite{[27]} described by him with ML algorithms to enable the machines to draw causal conclusions and simulate interventions.

The integration of causal methods in AI also contributes to increasing its transparency and thus the acceptance of AI methods, since a reference to probabilities or statistical correlations in the context of an explanation is not as effective as a reference to causes and causal effects \cite{[42]}.

\section{Evaluating uncertainty, interpretability and validation} \label{uncertainty}

Uncertainty quantification is often neglected in AI applications. One reason may be the above discussed misconception that `Big Data' automatically leads to exact results, making uncertainty quantification redundant. Another key reason is the complexity of the methods which hampers the construction of statistically valid uncertainty regions. However, most statisticians would agree that any comprehensive data analysis should contain methods to quantify the uncertainty of estimates and predictions. Its importance is also stressed by the American statistician David B. Dunson who writes that: \emph{`it is crucial to not over-state the results and appropriately characterize the (often immense) uncertainty to avoid flooding the scientific literature with false findings.'} \cite{[44]}

In fact, in order to achieve the main goal of highly accurate predictions, assumptions about underlying distributions and functional relationships are deliberately dropped in AI applications. On the one hand, this allows for a greater flexibility of the procedures. On the other hand, however, this also complicates an accurate quantification of uncertainty, e.g., to specify valid prediction and confidence regions for target variables and parameters of interest. As B\"uhlmann and colleagues put it: \emph{`The statistical theory serves as guard against cheating with data: you cannot beat the uncertainty principle.'} \cite{[45]}
In recent years, proposals for uncertainty quantification in AI methods have already been developed (by combinations with Bayesian approximations, bootstrapping, jackknifing and other cross-validation techniques, Gaussian processes, Monte Carlo dropout etc., see e.~g., \cite{[111],[112],[113],[120],[110]}). However, their theoretical validity (e.g., that a prediction interval actually covers future values 95\% of the time) has either not been demonstrate yet or only been proven under very restrictive or at least partially unrealistic assumptions.

In contrast, algorithmic methods could be embedded in statistical models. While potentially less flexible, they permit a better quantification of the underlying uncertainty by specifying valid prediction and confidence intervals. Thus, they allow for a better interpretation of the results. 

In comparison, the estimated parameters of many AI approaches (such as deep learning) are difficult to interpret. Pioneering work from computer science on this topic is, for example, \cite{[138],[139]}, for which Leslie Valiant was awarded the Turing Award in 2010. Further research is nevertheless needed to improve interpretability. This also includes uncertainty quantification of patterns identified by an AI method, which heavily rely on statistical techniques. A tempting approach to achieve more interpretable AI methods is the use of auxiliary models. These are comparatively simple statistical models which, after adaptation of a deep learning approach, describe the most important patterns represented by the AI method and potentially can also be used to quantify uncertainty \cite{[92],[91],[93],[90]}. In fact, as in computational and statistical learning theory \cite{[105],[58],[57]}, statistical methods and AI learning approaches can (and should) complement each other. Another important aspect is the model complexity which can, e.g., be captured by entropies (such as VC dimensions) or compression barriers \cite{[59]}. These concepts as well as different forms of regularization \cite{[114],[115],[116]}, i.e. the restriction of the parameter space, allow to recognize or even to correct an overfitting of a learning procedure. Here, the application of complexity reducing concepts can be seen as a direct implementation of the Lex Parsimoniae principle and often increases the interpretability of resulting models \cite{[118],[117]}. In fact, regularization and complexity reducing concepts are an integral part of many AI methods. However, they are also basic principles of modern statistics, which were already proposed before their introduction to AI. Examples are given in connection with empirical Bayesian or shrinkage methods \cite{roever2020}. In addition to that, AI and statistics have numerous concepts in common which give rise to an exchange of methods in these fields.

Beyond interpretability and uncertainty quantification, the above-mentioned validation aspects are of immense importance. Often in AI, validation is only carried out on single, frequently used `established' data sets. Thus, a certain stability or variability of the results cannot be reliably assessed due to the lack of generalizability. To tackle this problem we can again turn to statistical concepts: In order to reflect a diversity of real life situations, statistics makes use of probabilistic models. In addition to mathematical validity proofs and theoretical investigations, detailed simulation studies are carried out to evaluate the methods' limits (by exceeding the assumptions made). This statistical perspective provides extremely useful insights. Furthermore, validation aspects also apply to quality criteria (e.g., accuracy, sensitivity and specificity) of AI algorithms. The corresponding estimators are also random but their uncertainty is usually not quantified at all.

A particular challenge is posed by the ever faster development cycles of AI systems which need ongoing investigations on their adequate validation. This can even be aggravated when turning to development processes of mobile apps or online learning systems such as Amazon's recommender systems. Here, the developments are dynamic, de facto never ending processes, which therefore require continuous validation.

Statistics can help to increase the validity and interpretability of AI methods by providing contributions to the quantification of uncertainty. To achieve this, we can assume specific probabilistic and statistical models or dependency structures which allow comprehensive mathematical investigations \cite{[49],[46],[50],[105],[47],[48],ramosaj2019a}, e.g., by investigating robustness properties, proving asymptotic consistency or (finite) error bounds. On the other hand this also includes the preparation of (stochastic) simulation designs \cite{[51]} and the specification of easy to interpret auxiliary statistical models. Finally, it allows for a detailed analysis of quality criteria of AI algorithms.

\section{Education and training} \label{edu}

AI has been a growing research area for years, and its development is far from complete. In addition to ethical and legal problems, it has been shown that there are still many open questions regarding the collection and processing of data. Therefore, statistical methods must be considered as integral part of AI systems, from the formulation of the research questions, the development of the research design, through the analysis up to the interpretation of the results. Particularly in the field of methodological development, statistics can, e.g., serve as multiplier and strengthen the scientific exchange by establishing broad and strongly interconnected networks between users and developers. With its unique knowledge, statistics is a natural partner for other disciplines in teaching, research and practice.

{\bf School education:} In modern contemporary societies, the systematic inclusion of AI and its underlying statistical framework is --adapted to the cognitive abilities of the respective target populations-- essential at all levels of the educational system. Statistics and computer science should be integral elements of the curricula at schools. This would ensure that AI can be taught appropriately for the age in the different types of schools (primary, secondary, and vocational schools). For this purpose, school-related projects and teacher training initiatives must be initialized and promoted under the scientific supervision and based on the didactics of statistics and computer science. The goal is making the complex contents attractive and interesting for pupils by taking into account technical and social aspects. Pilot projects are, for example, the `Project Data Science and Big Data in Schools' (PRODABI; www.prodabi.de) \cite{[137]}, the `Introduction to Data Science' project (IDS; www.introdatascience.org), and the `International Data Science in Schools Project' (IDSSP; www.idssp.org). Particularly the latter two international projects have a pioneering role in school-related projects of stimulating AI literacy. To sum up, statistics should contribute its expertise to a greater extent to school curricula to ensure a broad and sustainable increase in statistical and computer literacy as a key resource for the future development of seminal technologies. Furthermore, teachers particularly qualified in statistics and its didactics should be systematically involved in digital teaching and training offers (e.g., e-learning).

{\bf Higher education:} For the tertiary sector, it is important to provide all relevant disciplines with the scientific foundations for high-level basic and applied research in the context of AI. This applies to both, Data Science subjects, such as such as mathematics, statistics and computer science, as well as corss-sectional areas, such as medicine, engineering, social and economic sciences. This includes adapting staffing policies at the universities as well as funding programs, possibly new courses of study, doctoral programs, research associations and research programs. In this regard, we see a particular need for expanding the qualification of teachers in statistical methodology and didactics due to the growing demand.

{\bf Training and interdisciplinary exchange}: Continuing professional development and training should be expanded at various levels. Here, advanced training programs on AI in various forms and formats are conceivable: Informatics and Engineering departments of various universities already offer workshops/summer schools on AI, for example the AI-DLDA at the Universita degli studi di Udine, Italy, \url{https://www.ip4fvg.it/summer-school/} or the AI Summer School in Singapore, \url{https://aisummerschool.aisingapore.org/}.
Professional development programs in this context include, for example, Coursera (\url{https://www.coursera.org/}), which offers online courses on various topics in collaboration with more than 200 leading universities and companies worldwide. Other possibilities include webinars, mentoring, laboratory visits, etc. Similar examples already exist for other areas of statistics such as Biometrics or Medical Informatics, see, for example, the certificates provided by the German Association for Medical Informatics, Biometry and Epidemiology (GMDS).
It is important to expand existing offers and establish new ones with a special focus on the statistical contributions to AI. In particular, training should be offered for both methodologists such as computer scientists, statisticians, and mathematicians who are not yet working in the field of AI as well as for users such as clinicians, engineers, social scientists and economists. 

By developing professional networks, participating methodologists can be brought together with users/experts to establish or maintain a continuous exchange between the disciplines. In addition to AI methods, these events should also cover the topics of data curation, management of data quality and data integration.

\section{Conclusions} \label{summary}

Statistics is a broad cross-scientific discipline. It provides knowledge and experience of all aspects of data evaluation: starting with the research question through design and analysis to the interpretation. As a core element of AI, it is the natural partner for other disciplines in teaching, research and practice. In particular, the following contributions of statistics to the field of artificial intelligence can be summarized:

\begin{enumerate}
\item Methodological development: The development of AI systems and their theoretical underpinning has benefited greatly from research in computer science and statistics, and many procedures have been developed by statisticians. Recent advances such as extreme learning machines show that statistics also provides important contributions to the design of AI systems, for example, by improved learning algorithms based on penalized or robust estimation methods.
\item Planning and design: Statistics can help to optimize data collection or preparation (sample size, sampling design, weighting, restriction of the data set, design of experiments, etc.) for subsequent evaluation with AI methods. Furthermore, the quality measures of statistics and their associated inference methods can help in the evaluation of AI models.
\item Assessment of data quality and data collection: Exploratory data analysis provides a wide range of tools to visualize the empirical distribution of the data and to derive appropriate metrics, which can be used to detect anomalies or to define ranges of typical values, to correct input errors, to determine norm values and to impute missing values. In combination with standardization in data storage, errors in the measurement process can be detected and corrected at an early stage. With the help of model-based statistical methods, comprehensive parameter tuning is also possible, even for small data sets.
\item Differentiation of causality and associations: In statistics, methods for dealing with covariate effects are known. Here, it is important to differentiate theoretically informed between the different relationships covariates can have to treatment and outcome in order to avoid bias in the estimation of causal effects. Pearl's causal framework enables the analysis of causal effects and the simulation of interventions. The integration of causal methods into AI can also contribute to the transparency and acceptance of AI methods.
\item Assessment of certainty or uncertainty in results: Statistics can help to enable or improve the quantification of uncertainty in and the interpretability of AI methods. By adopting specific statistical models, mathematical proofs of validity can also be provided. In addition, limitations of the methods can be explored through (stochastic) simulation designs.
\item Conscientious implementation of points 2 to 5, including a previously defined evaluation plan, also counteracts the replication crisis \cite{[5]} in many scientific disciplines. In this methodological crisis, which has been ongoing since the beginning of the 2010s, it has become clear that many studies, especially in medicine and the social sciences, are difficult or impossible to reproduce. 
\item Education, advanced vocational training and public relations: With its specialized knowledge, statistics is the natural partner for other disciplines in teaching and training. Especially in the further development of methods of artificial intelligence, statistics can strengthen scientific exchange.
\end{enumerate}

The objective of statistics related to AI must be to facilitate the interpretation of data. Data alone is hardly a science, no matter how great their volume or how subtly they are manipulated. What is important is the knowledge gained that will enable future interventions \cite{[43]}.

\section*{Declarations}

\subsection*{Funding}

Not applicable

\subsection*{Conflict of interest}

The authors declare that they have no conflict of interest.

\subsection*{Availability of data and material (data transparency)}

Not applicable

\subsection*{Code availability (software application or custom code)}

Not applicable

\begin{acknowledgements}
We would like to thank Rolf Biehler for his valuable input on Data Science projects at schools. 

This paper is based on the position paper `DAGStat Stellungnahme: Die Rolle der Statistik in der K\"unstlichen Intelligenz` (\url{https://www.dagstat.de/fileadmin/dagstat/documents/DAGStat_KI_Stellungnahme_200303.pdf}), which has been drafted by a working group of the German Consortium in Statistics (DAGStat) and approved by the members.

\end{acknowledgements}

%


\bibliographystyle{spmpsci}      
\bibliography{Lit}   

\begin{thebibliography}{100}
\providecommand{\url}[1]{{#1}}
\providecommand{\urlprefix}{URL }
\expandafter\ifx\csname urlstyle\endcsname\relax
  \providecommand{\doi}[1]{DOI~\discretionary{}{}{}#1}\else
  \providecommand{\doi}{DOI~\discretionary{}{}{}\begingroup
  \urlstyle{rm}\Url}\fi

\bibitem{[85]}
AInow: \url{https://ainowinstitute.org/}.
\newblock Accessed 02.02.2020

\bibitem{[40]}
Athey, S., Imbens, G.W.: Machine learning for estimating heterogeneous causal
  effects.
\newblock Tech. rep., Stanford University, Graduate School of Business (2015)

\bibitem{[38]}
Athey, S., Imbens, G.W.: The econometrics of randomized experiments.
\newblock In: Handbook of Economic Field Experiments, vol.~1, pp. 73--140.
  Elsevier (2017)

\bibitem{[49]}
Athey, S., Tibshirani, J., Wager, S., et~al.: Generalized random forests.
\newblock The Annals of Statistics \textbf{47}(2), 1148--1178 (2019)

\bibitem{[100]}
Barrachina, S., Bender, O., Casacuberta, F., Civera, J., Cubel, E., Khadivi,
  S., Lagarda, A., Ney, H., Tom{\'{a}}s, J., Vidal, E., Vilar, J.M.:
  Statistical approaches to computer-assisted translation.
\newblock Computational Linguistics \textbf{35}(1), 3--28 (2009).
\newblock \doi{10.1162/coli.2008.07-055-r2-06-29}.
\newblock \urlprefix\url{https://doi.org/10.1162%2Fcoli.2008.07-055-r2-06-29}

\bibitem{[125]}
Bartels, D.M., Hastie, R., Urminsky, O.: Connecting laboratory and field
  research in judgment and decision making: Causality and the breadth of
  external validity.
\newblock Journal of Applied Research in Memory and Cognition \textbf{7}(1),
  11--15 (2018).
\newblock \doi{10.1016/j.jarmac.2018.01.001}.
\newblock \urlprefix\url{https://doi.org/10.1016%2Fj.jarmac.2018.01.001}

\bibitem{[46]}
Bartlett, P.L., Bickel, P.J., B{\"u}hlmann, P., Freund, Y., Friedman, J.,
  Hastie, T., Jiang, W., Jordan, M.J., Koltchinskii, V., Lugosi, G., et~al.:
  Discussions of boosting papers, and rejoinders.
\newblock The Annals of Statistics \textbf{32}(1), 85--134 (2004)

\bibitem{[4]}
Bellman, R.: Dynamic programming.
\newblock Princeton University Press (1957)

\bibitem{[137]}
Biehler, R., Budde, L., Frischemeier, D., Heinemann, B., Podworny, S., Schulte,
  C., Wassong, T. (eds.): Paderborn Symposium on Data Science Education at
  School Level 2017: The Collected Extended Abstracts.
\newblock Paderborn: Universitätsbibliothek Paderborn (2018).
\newblock \url{http://dx.doi.org/10.17619/UNIPB/1-374}

\bibitem{Brantingham2018}
Brantingham, P.J., Valasik, M., Mohler, G.O.: {Does predictive policing lead to
  biased arrests? Results from a randomized controlled trial}.
\newblock Statistics and public policy \textbf{5}(1), 1--6 (2018)

\bibitem{[126]}
Braver, S.L., Smith, M.C.: Maximizing both external and internal validity in
  longitudinal true experiments with voluntary treatments: The
  {\textquotedblleft}combined modified{\textquotedblright} design.
\newblock Evaluation and Program Planning \textbf{19}(4), 287--300 (1996).
\newblock \doi{10.1016/s0149-7189(96)00029-8}.
\newblock \urlprefix\url{https://doi.org/10.1016%2Fs0149-7189%2896%2900029-8}

\bibitem{[98]}
Breiman, L.: Bagging predictors.
\newblock Machine Learning \textbf{24}(2), 123--140 (1996).
\newblock \doi{10.1007/bf00058655}.
\newblock \urlprefix\url{https://doi.org/10.1007%2Fbf00058655}

\bibitem{[82]}
Breiman, L.: Random forests.
\newblock Machine learning \textbf{45}(1), 5--32 (2001)

\bibitem{[45]}
B{\"u}hlmann, P., van~de Geer, S.: Statistics for big data: A perspective.
\newblock Statistics \& Probability Letters \textbf{136}, 37--41 (2018)

\bibitem{[103]}
{Bundespolizeipr\"asidium Potsdam}: {Abschlussbericht Teilprojekt 1
  "Biometrische Gesichtserkennung"}.
\newblock
  \url{https://www.bundespolizei.de/Web/DE/04Aktuelles/01Meldungen/2018/10/181011_abschlussbericht_gesichtserkennung_down.pdf?__blob=publicationFile=1}
  (2018).
\newblock Accessed 07.05.2020

\bibitem{[1]}
Bundesregierung: Artificial intelligence strategy.
\newblock
  \url{https://www.ki-strategie-deutschland.de/home.html?file=files/downloads/Nationale_KI-Strategie_engl.pdf}
  (2018).
\newblock Accessed 07.05.2020

\bibitem{[23]}
Burt, J.R., Torosdagli, N., Khosravan, N., RaviPrakash, H., Mortazi, A.,
  Tissavirasingham, F., Hussein, S., Bagci, U.: Deep learning beyond cats and
  dogs: recent advances in diagnosing breast cancer with deep neural networks.
\newblock The British journal of radiology \textbf{91}(1089), 20170545 (2018)

\bibitem{[123]}
{Catalogue of bias collaboration, Lee H, Aronson JK, Nunan D}: Catalogue of
  bias: Collider bias.
\newblock \url{https://catalogofbias.org/biases/collider-bias} (2019).
\newblock Accessed 12.02.2020

\bibitem{[104]}
Chen, C.L.P., Liu, Z.: Broad learning system: An effective and efficient
  incremental learning system without the need for deep architecture.
\newblock {IEEE} Transactions on Neural Networks and Learning Systems
  \textbf{29}(1), 10--24 (2018).
\newblock \doi{10.1109/tnnls.2017.2716952}.
\newblock \urlprefix\url{https://doi.org/10.1109%2Ftnnls.2017.2716952}

\bibitem{[24]}
Chen, H., Engkvist, O., Wang, Y., Olivecrona, M., Blaschke, T.: The rise of
  deep learning in drug discovery.
\newblock Drug discovery today \textbf{23}(6), 1241--1250 (2018)

\bibitem{[106]}
Chen, S., Cowan, C., Grant, P.: Orthogonal least squares learning algorithm for
  radial basis function networks.
\newblock {IEEE} Transactions on Neural Networks \textbf{2}(2), 302--309
  (1991).
\newblock \doi{10.1109/72.80341}.
\newblock \urlprefix\url{https://doi.org/10.1109%2F72.80341}

\bibitem{[34]}
Cochran, W.G., Rubin, D.B.: Controlling bias in observational studies: A
  review.
\newblock Sankhy{\=a}: The Indian Journal of Statistics, Series A pp. 417--446
  (1973)

\bibitem{[81]}
Cortes, C., Vapnik, V.: Support-vector networks.
\newblock Machine Learning \textbf{20}(3), 273--297 (1995).
\newblock \doi{10.1007/bf00994018}.
\newblock \urlprefix\url{https://doi.org/10.1007%2Fbf00994018}

\bibitem{[20]}
Dastin, J.: {Amazon scraps secret AI recruiting tool that showed bias against
  women. Reuters (2018)}.
\newblock
  \url{https://www.reuters.com/article/us-amazon-com-jobs-automation-insight/amazon-scraps-secret-ai-recruiting-tool-that-showed-bias-against-women-idUSKCN1MK08G}
  (2018).
\newblock Accessed 27.11.2019

\bibitem{[107]}
{Data Ethics Commission of the Federal Government, Federal Ministry of the
  Interior, Building and Community}: Opinion of the data ethics commission.
\newblock
  \url{https://www.bmi.bund.de/SharedDocs/downloads/EN/themen/it-digital-policy/datenethikkommission-abschlussgutachten-lang.pdf?__blob=publicationFile&v=4}
  (2019).
\newblock Accessed 07.05.2020

\bibitem{[12]}
Davis, E.: {AI amusements: the tragic tale of Tay the chatbot}.
\newblock AI Matters \textbf{2}(4), 20--24 (2016)

\bibitem{[50]}
Devroye, L., Gy{\"o}rfi, L., Lugosi, G.: A probabilistic theory of pattern
  recognition, vol.~31.
\newblock Springer Science \& Business Media (2013)

\bibitem{[36]}
Didelez, V.: Graphical models for composable finite markov processes.
\newblock Scandinavian Journal of Statistics \textbf{34}(1), 169--185 (2007)

\bibitem{[37]}
Duflo, E., Glennerster, R., Kremer, M.: Using randomization in development
  economics research: A toolkit.
\newblock Handbook of development economics \textbf{4}, 3895--3962 (2007)

\bibitem{[61]}
Duke-Margolis:
  \url{https://healthpolicy.duke.edu/sites/default/files/atoms/files/characterizing_rwd.pdf}
  (2018).
\newblock Accessed 13.05.2020

\bibitem{[62]}
Duke-Margolis:
  \url{https://healthpolicy.duke.edu/sites/default/files/u31/rwd_reliability.pdf}
  (2019).
\newblock Accessed 13.05.2020

\bibitem{[44]}
Dunson, D.B.: Statistics in the big data era: Failures of the machine.
\newblock Statistics \& Probability Letters \textbf{136}, 4--9 (2018)

\bibitem{[89]}
Einav, L., Levin, J.: The data revolution and economic analysis.
\newblock Innovation Policy and the Economy \textbf{14}, 1--24 (2014).
\newblock \doi{10.1086/674019}.
\newblock \urlprefix\url{https://doi.org/10.1086%2F674019}

\bibitem{[102]}
{European Commission}:
  \url{https://ec.europa.eu/info/resources-partners/machine-translation-public-administrations-etranslation_en#translateonline}
  (2020).
\newblock Accessed 13.05.2020

\bibitem{europ}
{European Commission}: {On Artificial Intelligence - A European approach to
  excellence and trust}.
\newblock
  \url{https://ec.europa.eu/info/sites/info/files/commission-white-paper-artificial-intelligence-feb2020_en.pdf}
  (2020).
\newblock Accessed 29.07.2020

\bibitem{[128]}
{European Statistical System}: Quality assurance framework of the european
  statistical system.
\newblock
  \url{https://ec.europa.eu/eurostat/documents/64157/4392716/ESS-QAF-V1-2final.pdf/bbf5970c-1adf-46c8-afc3-58ce177a0646}
  (2019).
\newblock Accessed 07.05.2020

\bibitem{[74]}
Fakoor, R., Ladhak, F., Nazi, A., Huber, M.: Using deep learning to enhance
  cancer diagnosis and classification.
\newblock In: Proceedings of the international conference on machine learning,
  vol.~28. ACM New York, USA (2013)

\bibitem{[121]}
Fayyad, U., Piatetsky-Shapiro, G., Smyth, P.: From data mining to knowledge
  discovery in databases.
\newblock AI magazine \textbf{17}(3), 37--37 (1996)

\bibitem{[7]}
FDA: \url{https://www.fda.gov/media/122535/download} (2019).
\newblock Accessed 13.05.2020

\bibitem{[66]}
Feng, Q.Y., Vasile, R., Segond, M., Gozolchiani, A., Wang, Y., Abel, M.,
  Havlin, S., Bunde, A., Dijkstra, H.A.: {ClimateLearn}: A machine-learning
  approach for climate prediction using network measures.
\newblock Geoscientific Model Development Discussions pp. 1--18 (2016).
\newblock \doi{10.5194/gmd-2015-273}.
\newblock \urlprefix\url{https://doi.org/10.5194%2Fgmd-2015-273}

\bibitem{[142]}
{Forbes}:
  \url{https://www.forbes.com/sites/bernardmarr/2018/03/05/heres-why-data-is-not-the-new-oil/#45b487143aa9}
  (2018).
\newblock Accessed 27.04.2020

\bibitem{[73]}
Foster, K.R., Koprowski, R., Skufca, J.D.: Machine learning, medical diagnosis,
  and biomedical engineering research - commentary.
\newblock {BioMedical} Engineering {OnLine} \textbf{13}(1), 94 (2014).
\newblock \doi{10.1186/1475-925x-13-94}.
\newblock \urlprefix\url{https://doi.org/10.1186%2F1475-925x-13-94}

\bibitem{[97]}
Freund, Y., Schapire, R.E.: A decision-theoretic generalization of on-line
  learning and an application to boosting.
\newblock Journal of Computer and System Sciences \textbf{55}(1), 119--139
  (1997).
\newblock \doi{10.1006/jcss.1997.1504}.
\newblock \urlprefix\url{https://doi.org/10.1006%2Fjcss.1997.1504}

\bibitem{[86]}
Friedman, J., Hastie, T., Tibshirani, R.: The elements of statistical learning.
\newblock Springer series in statistics New York (2001)

\bibitem{[63]}
Gabler, S., H\"ader, S.: {Repr\"asentativit\"at: Versuch einer
  Begriffsbestimmung}.
\newblock In: Telefonumfragen in Deutschland, pp. 81--112. Springer Fachmedien
  Wiesbaden (2018).
\newblock \doi{10.1007/978-3-658-23950-3_5}.
\newblock \urlprefix\url{https://doi.org/10.1007%2F978-3-658-23950-3_5}

\bibitem{[111]}
Gal, Y., Ghahramani, Z.: Dropout as a bayesian approximation: Representing
  model uncertainty in deep learning.
\newblock In: international conference on machine learning, pp. 1050--1059
  (2016)

\bibitem{[112]}
Garnelo, M., Rosenbaum, D., Maddison, C.J., Ramalho, T., Saxton, D., Shanahan,
  M., Teh, Y.W., Rezende, D.J., Eslami, S.: Conditional neural processes.
\newblock arXiv preprint arXiv:1807.01613  (2018)

\bibitem{[31]}
Greenland, S., Robins, J.M.: Identifiability, exchangeability, and
  epidemiological confounding.
\newblock International journal of epidemiology \textbf{15}(3), 413--419 (1986)

\bibitem{[32]}
Greenland, S., Robins, J.M., Pearl, J., et~al.: Confounding and collapsibility
  in causal inference.
\newblock Statistical science \textbf{14}(1), 29--46 (1999)

\bibitem{gudivada2015}
Gudivada, V.N., Baeza-Yates, R., Raghavan, V.V.: Big data: Promises and
  problems.
\newblock Computer \textbf{48}(3), 20--23 (2015).
\newblock \doi{10.1109/MC.2015.62}

\bibitem{[105]}
Gy{\"o}rfi, L., Kohler, M., Krzyzak, A., Walk, H.: A Distribution-Free Theory
  of Nonparametric Regression.
\newblock Springer New York (2002).
\newblock \doi{10.1007/b97848}.
\newblock \urlprefix\url{https://doi.org/10.1007%2Fb97848}

\bibitem{[135]}
Heckman, J.J.: Micro data, heterogeneity, and the evaluation of public policy:
  Nobel lecture.
\newblock Journal of Political Economy \textbf{109}(4), 673--748 (2001).
\newblock \doi{10.1086/322086}.
\newblock \urlprefix\url{https://doi.org/10.1086%2F322086}

\bibitem{[16]}
Higgins, J.P., Altman, D.G., G{\o}tzsche, P.C., J{\"u}ni, P., Moher, D., Oxman,
  A.D., Savovi{\'c}, J., Schulz, K.F., Weeks, L., Sterne, J.A.: The cochrane
  collaboration’s tool for assessing risk of bias in randomised trials.
\newblock Bmj \textbf{343}, d5928 (2011)

\bibitem{[28]}
Hill, A.B.: The environment and disease: association or causation? (1965)

\bibitem{[69]}
Hsich, E., Gorodeski, E.Z., Blackstone, E.H., Ishwaran, H., Lauer, M.S.:
  Identifying important risk factors for survival in patient with systolic
  heart failure using random survival forests.
\newblock Circulation: Cardiovascular Quality and Outcomes \textbf{4}(1),
  39--45 (2011)

\bibitem{[99]}
Huang, G.B., Zhu, Q.Y., Siew, C.K.: Extreme learning machine: Theory and
  applications.
\newblock Neurocomputing \textbf{70}(1-3), 489--501 (2006).
\newblock \doi{10.1016/j.neucom.2005.12.126}.
\newblock \urlprefix\url{https://doi.org/10.1016%2Fj.neucom.2005.12.126}

\bibitem{[87]}
James, G., Witten, D., Hastie, T., Tibshirani, R.: An Introduction to
  Statistical Learning.
\newblock Springer New York (2013).
\newblock \doi{10.1007/978-1-4614-7138-7}.
\newblock \urlprefix\url{https://doi.org/10.1007%2F978-1-4614-7138-7}

\bibitem{[94]}
Juang, B.H., Rabiner, L.R.: Hidden markov models for speech recognition.
\newblock Technometrics \textbf{33}(3), 251--272 (1991).
\newblock \doi{10.1080/00401706.1991.10484833}.
\newblock \urlprefix\url{https://doi.org/10.1080%2F00401706.1991.10484833}

\bibitem{[25]}
Kamilaris, A., Prenafeta-Bold{\'u}, F.X.: Deep learning in agriculture: A
  survey.
\newblock Computers and electronics in agriculture \textbf{147}, 70--90 (2018)

\bibitem{karlin1966optimal}
Karlin, S., Studden, W.J.: Optimal experimental designs.
\newblock The Annals of Mathematical Statistics \textbf{37}(4), 783--815 (1966)

\bibitem{[19]}
Karr, A.F., Sanil, A.P., Banks, D.L.: Data quality: A statistical perspective.
\newblock Statistical Methodology \textbf{3}(2), 137--173 (2006)

\bibitem{[58]}
Kearns, M.J., Vazirani, U.: An Introduction to Computational Learning Theory.
\newblock The {MIT} Press (1994).
\newblock \doi{10.7551/mitpress/3897.001.0001}.
\newblock \urlprefix\url{https://doi.org/10.7551%2Fmitpress%2F3897.001.0001}

\bibitem{[78]}
Kelley, H.J.: Gradient theory of optimal flight paths.
\newblock {ARS} Journal \textbf{30}(10), 947--954 (1960).
\newblock \doi{10.2514/8.5282}.
\newblock \urlprefix\url{https://doi.org/10.2514%2F8.5282}

\bibitem{[95]}
Koch, C.: How the computer beat the go player.
\newblock Scientific American Mind \textbf{27}(4), 20--23 (2016).
\newblock \doi{10.1038/scientificamericanmind0716-20}.
\newblock
  \urlprefix\url{https://doi.org/10.1038%2Fscientificamericanmind0716-20}

\bibitem{[119]}
Kohavi, R., Tang, D., Xu, Y., Hemkens, L.G., Ioannidis, J.P.A.: Online
  randomized controlled experiments at scale: lessons and extensions to
  medicine.
\newblock Trials \textbf{21}(1) (2020).
\newblock \doi{10.1186/s13063-020-4084-y}.
\newblock \urlprefix\url{https://doi.org/10.1186%2Fs13063-020-4084-y}

\bibitem{[101]}
Kozielski, M., Doetsch, P., Ney, H.: Improvements in {RWTH}'s system for
  off-line handwriting recognition.
\newblock In: 2013 12th International Conference on Document Analysis and
  Recognition. {IEEE} (2013).
\newblock \doi{10.1109/icdar.2013.190}.
\newblock \urlprefix\url{https://doi.org/10.1109%2Ficdar.2013.190}

\bibitem{kruskal1}
Kruskal, W., Mosteller, F.: {Representative sampling, I: non-scientific
  literature}.
\newblock International Statistical Review/Revue Internationale de Statistique
  pp. 13--24 (1979)

\bibitem{kruskal2}
Kruskal, W., Mosteller, F.: {Representative sampling, II: Scientific
  literature, excluding statistics}.
\newblock International Statistical Review/Revue Internationale de Statistique
  pp. 111--127 (1979)

\bibitem{kruskal3}
Kruskal, W., Mosteller, F.: {Representative sampling, III: The current
  statistical literature}.
\newblock International Statistical Review/Revue Internationale de Statistique
  pp. 245--265 (1979)

\bibitem{kruskal4}
Kruskal, W., Mosteller, F.: {Representative sampling, IV: The history of the
  concept in statistics, 1895-1939}.
\newblock International Statistical Review/Revue Internationale de Statistique
  pp. 169--195 (1980)

\bibitem{[39]}
Van~der Laan, M.J., Rose, S.: Targeted learning: causal inference for
  observational and experimental data.
\newblock Springer Science \& Business Media (2011)

\bibitem{[59]}
Langford, J.: Tutorial on practical prediction theory for classification.
\newblock Journal of machine learning research \textbf{6}(Mar), 273--306 (2005)

\bibitem{[11]}
Lazer, D., Kennedy, R., King, G., Vespignani, A.: {The parable of Google Flu:
  traps in big data analysis}.
\newblock Science \textbf{343}(6176), 1203--1205 (2014)

\bibitem{[70]}
Lee, B.J., Kim, J.Y.: Identification of type 2 diabetes risk factors using
  phenotypes consisting of anthropometry and triglycerides based on machine
  learning.
\newblock {IEEE} Journal of Biomedical and Health Informatics \textbf{20}(1),
  39--46 (2016).
\newblock \doi{10.1109/jbhi.2015.2396520}.
\newblock \urlprefix\url{https://doi.org/10.1109%2Fjbhi.2015.2396520}

\bibitem{[21]}
Levine, S., Pastor, P., Krizhevsky, A., Ibarz, J., Quillen, D.: Learning
  hand-eye coordination for robotic grasping with deep learning and large-scale
  data collection.
\newblock The International Journal of Robotics Research \textbf{37}(4-5),
  421--436 (2018)

\bibitem{[71]}
Li, Y., Gong, S., Liddell, H.: Support vector regression and classification
  based multi-view face detection and recognition.
\newblock In: Proceedings Fourth {IEEE} International Conference on Automatic
  Face and Gesture Recognition (Cat. No. {PR}00580). {IEEE} Comput. Soc (2000).
\newblock \doi{10.1109/afgr.2000.840650}.
\newblock \urlprefix\url{https://doi.org/10.1109%2Fafgr.2000.840650}

\bibitem{[88]}
Libbrecht, M.W., Noble, W.S.: Machine learning applications in genetics and
  genomics.
\newblock Nature Reviews Genetics \textbf{16}(6), 321--332 (2015).
\newblock \doi{10.1038/nrg3920}.
\newblock \urlprefix\url{https://doi.org/10.1038%2Fnrg3920}

\bibitem{[60]}
Lin, E.J.D., Hefner, J.L., Zeng, X., Moosavinasab, S., Huber, T., Klima, J.,
  Liu, C., Lin, S.M.: Currently reading a deep learning model for pediatric
  patient risk stratification.
\newblock The American Journal of Managed Care  (2019)

\bibitem{[129]}
McCracken, M.W., Ng, S.: {FRED}-{MD}: A monthly database for macroeconomic
  research.
\newblock Journal of Business {\&} Economic Statistics \textbf{34}(4), 574--589
  (2016).
\newblock \doi{10.1080/07350015.2015.1086655}.
\newblock \urlprefix\url{https://doi.org/10.1080%2F07350015.2015.1086655}

\bibitem{[84]}
MedTechIntelligence:
  \url{https://www.medtechintelligence.com/news_article/apple-watch-4-gets-fda-clearance/}
  (2018).
\newblock Accessed 13.05.2020

\bibitem{[9]}
Meng, X.L.: {Statistical paradises and paradoxes in big data (I): Law of large
  populations, big data paradox, and the 2016 US presidential election}.
\newblock The Annals of Applied Statistics \textbf{12}(2), 685--726 (2018)

\bibitem{[10]}
Meng, X.L., Xie, X.: {I got more data, my model is more refined, but my
  estimator is getting worse! Am I just dumb?}
\newblock Econometric Reviews \textbf{33}(1-4), 218--250 (2014)

\bibitem{[42]}
Miller, T.: Explanation in artificial intelligence: Insights from the social
  sciences.
\newblock Artificial Intelligence \textbf{267}, 1--38 (2019)

\bibitem{[108]}
Molenberghs, G., Fitzmaurice, G., Kenward, M.G., Tsiatis, A. (eds.): Handbook
  of Missing Data Methodology.
\newblock Chapman and Hall/{CRC} (2014).
\newblock \doi{10.1201/b17622}.
\newblock \urlprefix\url{https://doi.org/10.1201%2Fb17622}

\bibitem{[92]}
Molnar, C.: Interpretable machine learning.
\newblock \url{https://christophm. github. io/interpretable-ml-book/} pp.
  05--04 (2019).
\newblock Accessed 29.07.2020

\bibitem{[54]}
Moor, J.: The dartmouth college artificial intelligence conference: The next
  fifty years.
\newblock Ai Magazine \textbf{27}(4), 87--87 (2006)

\bibitem{[51]}
Morris, T.P., White, I.R., Crowther, M.J.: Using simulation studies to evaluate
  statistical methods.
\newblock Statistics in medicine \textbf{38}(11), 2074--2102 (2019)

\bibitem{[141]}
{New York Times}:
  \url{https://www.nytimes.com/2018/12/18/technology/facebook-privacy.html}
  (2018).
\newblock Accessed 27.04.2020

\bibitem{[130]}
Ng, S.: Opportunities and challenges: Lessons from analyzing terabytes of
  scanner data.
\newblock In: B.~Honore, A.~Pakes, M.~Piazzesi, L.~Samuelson (eds.) Advances in
  Economics and Econometrics, pp. 1--34. Cambridge University Press (2018).
\newblock \doi{10.1017/9781108227223.001}.
\newblock \urlprefix\url{https://doi.org/10.1017%2F9781108227223.001}

\bibitem{[65]}
Ortega, H., Li, H., Suruki, R., Albers, F., Gordon, D., Yancey, S.: Cluster
  analysis and characterization of response to mepolizumab. a step closer to
  personalized medicine for patients with severe asthma.
\newblock Annals of the American Thoracic Society \textbf{11}(7), 1011--1017
  (2014).
\newblock \doi{10.1513/annalsats.201312-454oc}.
\newblock \urlprefix\url{https://doi.org/10.1513%2Fannalsats.201312-454oc}

\bibitem{[113]}
Osband, I., Blundell, C., Pritzel, A., Van~Roy, B.: Deep exploration via
  bootstrapped dqn.
\newblock In: Advances in neural information processing systems, pp. 4026--4034
  (2016)

\bibitem{[5]}
Pashler, H., Wagenmakers, E.J.: Editors’ introduction to the special section
  on replicability in psychological science: A crisis of confidence?
\newblock Perspectives on Psychological Science \textbf{7}(6), 528--530 (2012)

\bibitem{[72]}
Patil, P.: Multilayered network for {LPC} based speech recognition.
\newblock {IEEE} Transactions on Consumer Electronics \textbf{44}(2), 435--438
  (1998).
\newblock \doi{10.1109/30.681960}.
\newblock \urlprefix\url{https://doi.org/10.1109%2F30.681960}

\bibitem{[80]}
Pearl, J.: Probabilistic reasoning in intelligent systems.
\newblock San Mateo, CA: Kaufmann \textbf{23}, 33--34 (1988)

\bibitem{[33]}
Pearl, J.: Aspects of graphical models connected with causality.
\newblock In: Proceedings of the 49th Session of the International Statistical
  Science Institute (1993)

\bibitem{[30]}
Pearl, J.: Causality.
\newblock Cambridge University Press (2009)

\bibitem{[27]}
Pearl, J.: The foundations of causal inference.
\newblock Sociological Methodology \textbf{40}(1), 75--149 (2010)

\bibitem{[43]}
Pearl, J.: Theoretical impediments to machine learning with seven sparks from
  the causal revolution.
\newblock arXiv preprint arXiv:1801.04016v1  (2018)

\bibitem{[91]}
Peltola, T.: Local interpretable model-agnostic explanations of bayesian
  predictive models via kullback-leibler projections.
\newblock arXiv preprint arXiv:1810.02678v1  (2018)

\bibitem{[14]}
Perez, M.V., Mahaffey, K.W., Hedlin, H., Rumsfeld, J.S., Garcia, A., Ferris,
  T., Balasubramanian, V., Russo, A.M., Rajmane, A., Cheung, L., et~al.:
  Large-scale assessment of a smartwatch to identify atrial fibrillation.
\newblock New England Journal of Medicine \textbf{381}(20), 1909--1917 (2019)

\bibitem{ramosaj2020}
Ramosaj, B., Amro, L., Pauly, M.: A cautionary tale on using imputation methods
  for inference in matched pairs design.
\newblock Bioinformatics \textbf{36}(10), 3099--3106 (2020)

\bibitem{ramosaj2019a}
Ramosaj, B., Pauly, M.: Consistent estimation of residual variance with random
  forest out-of-bag errors.
\newblock Statistics {\&} Probability Letters \textbf{151}, 49--57 (2019)

\bibitem{ramosaj2019}
Ramosaj, B., Pauly, M.: Predicting missing values: a comparative study on
  non-parametric approaches for imputation.
\newblock Computational Statistics \textbf{34}(4), 1741--1764 (2019)

\bibitem{[93]}
Ribeiro, M., Singh, S., Guestrin, C.: {{\textquotedblleft}Why Should I Trust
  You?{\textquotedblright}: Explaining the Predictions of Any Classifier}.
\newblock In: Proceedings of the 2016 Conference of the North American Chapter
  of the Association for Computational Linguistics: Demonstrations. Association
  for Computational Linguistics (2016).
\newblock \doi{10.18653/v1/n16-3020}.
\newblock \urlprefix\url{https://doi.org/10.18653%2Fv1%2Fn16-3020}

\bibitem{[90]}
Ribeiro, M.T., Singh, S., Guestrin, C.: Model-agnostic interpretability of
  machine learning.
\newblock arXiv preprint arXiv:1606.05386  (2016)

\bibitem{richter2019}
Richter, J., Madjar, K., Rahnenf{\"u}hrer, J.: Model-based optimization of
  subgroup weights for survival analysis.
\newblock Bioinformatics \textbf{35}(14), 484--491 (2019)

\bibitem{[35]}
Robins, J.M., Hernan, M.A., Brumback, B.: Marginal structural models and causal
  inference in epidemiology (2000)

\bibitem{[67]}
Robinson, C., Dilkina, B.: A machine learning approach to modeling human
  migration.
\newblock In: Proceedings of the 1st {ACM} {SIGCAS} Conference on Computing and
  Sustainable Societies, pp. 1--8. {ACM} Press (2018).
\newblock \doi{10.1145/3209811.3209868}.
\newblock \urlprefix\url{https://doi.org/10.1145%2F3209811.3209868}

\bibitem{[127]}
Roe, B.E., Just, D.R.: Internal and external validity in economics research:
  Tradeoffs between experiments, field experiments, natural experiments, and
  field data.
\newblock American Journal of Agricultural Economics \textbf{91}(5), 1266--1271
  (2009).
\newblock \doi{10.1111/j.1467-8276.2009.01295.x}.
\newblock \urlprefix\url{https://doi.org/10.1111%2Fj.1467-8276.2009.01295.x}

\bibitem{[132]}
Rosenbaum, P.: Observation and Experiment.
\newblock Harvard University Press (2017).
\newblock \doi{10.4159/9780674982697}.
\newblock \urlprefix\url{https://doi.org/10.4159%2F9780674982697}

\bibitem{[133]}
Rosenbaum, P.R.: Observational studies.
\newblock In: Springer Series in Statistics, pp. 1--17. Springer New York
  (2002).
\newblock \doi{10.1007/978-1-4757-3692-2_1}.
\newblock \urlprefix\url{https://doi.org/10.1007%2F978-1-4757-3692-2_1}

\bibitem{[134]}
Rosenbaum, P.R.: Design of Observational Studies.
\newblock Springer New York (2010).
\newblock \doi{10.1007/978-1-4419-1213-8}.
\newblock \urlprefix\url{https://doi.org/10.1007%2F978-1-4419-1213-8}

\bibitem{[77]}
Rosenblatt, F.: The perceptron: A probabilistic model for information storage
  and organization in the brain.
\newblock Psychological Review \textbf{65}(6), 386--408 (1958).
\newblock \doi{10.1037/h0042519}.
\newblock \urlprefix\url{https://doi.org/10.1037%2Fh0042519}

\bibitem{[118]}
Ross, A., Lage, I., Doshi-Velez, F.: The neural lasso: Local linear sparsity
  for interpretable explanations.
\newblock In: Workshop on Transparent and Interpretable Machine Learning in
  Safety Critical Environments, 31st Conference on Neural Information
  Processing Systems (2017)

\bibitem{roever2020}
R{\"o}ver, C., Friede, T.: Dynamically borrowing strength from another study
  through shrinkage estimation.
\newblock Statistical Methods in Medical Research \textbf{29}, 293--308 (2020)

\bibitem{[29]}
Rubin, D.B.: Estimating causal effects of treatments in randomized and
  nonrandomized studies.
\newblock Journal of educational Psychology \textbf{66}(5), 688 (1974)

\bibitem{rubin1976}
Rubin, D.B.: Inference and missing data.
\newblock Biometrika \textbf{63}(3), 581--592 (1976)

\bibitem{[131]}
Rubin, D.B.: Matched Sampling for Causal Effects.
\newblock Cambridge University Press (2006).
\newblock \doi{10.1017/cbo9780511810725}.
\newblock \urlprefix\url{https://doi.org/10.1017%2Fcbo9780511810725}

\bibitem{[17]}
Rubin, D.B.: For objective causal inference, design trumps analysis.
\newblock The Annals of Applied Statistics \textbf{2}(3), 808--840 (2008)

\bibitem{[55]}
Russell, S., Norvig, P.: Ai a modern approach.
\newblock Learning \textbf{2}(3), 4 (2005)

\bibitem{[83]}
Schmidhuber, J.: Deep learning in neural networks: An overview.
\newblock Neural Networks \textbf{61}, 85--117 (2015).
\newblock \doi{10.1016/j.neunet.2014.09.003}.
\newblock \urlprefix\url{https://doi.org/10.1016%2Fj.neunet.2014.09.003}

\bibitem{[47]}
Scornet, E., Biau, G., Vert, J.P., et~al.: Consistency of random forests.
\newblock The Annals of Statistics \textbf{43}(4), 1716--1741 (2015)

\bibitem{seaman2013}
Seaman, S.R., White, I.R.: Review of inverse probability weighting for dealing
  with missing data.
\newblock Statistical Methods in Medical Research \textbf{22}(3), 278--295
  (2013)

\bibitem{Searle1980}
Searle, J.: Minds, brains and programs.
\newblock Behavioral and Brain Science \textbf{3}(3), 417--457 (1980)

\bibitem{[76]}
Sese, A., Palmer, A.L., Montano, J.J.: Psychometric measurement models and
  artificial neural networks.
\newblock International Journal of Testing \textbf{4}(3), 253--266 (2004).
\newblock \doi{10.1207/s15327574ijt0403_4}.
\newblock \urlprefix\url{https://doi.org/10.1207%2Fs15327574ijt0403_4}

\bibitem{[8]}
Shadish, W.R., Cook, T.D., Campbell, D.T.: Experimental and quasi-experimental
  designs for generalized causal inference.
\newblock Boston: Houghton Mifflin (2002)

\bibitem{shah2014}
Shah, A.D., Bartlett, J.W., Carpenter, J., Nicholas, O., Hemingway, H.:
  Comparison of random forest and parametric imputation models for imputing
  missing data using mice: a caliber study.
\newblock American Journal of Epidemiology \textbf{179}(6), 764--774 (2014)

\bibitem{[68]}
Shen, S., Jiang, H., Zhang, T.: Stock market forecasting using machine learning
  algorithms.
\newblock Department of Electrical Engineering, Stanford University, Stanford,
  CA pp. 1--5 (2012)

\bibitem{[96]}
Silver, D., Hubert, T., Schrittwieser, J., Antonoglou, I., Lai, M., Guez, A.,
  Lanctot, M., Sifre, L., Kumaran, D., Graepel, T., Lillicrap, T., Simonyan,
  K., Hassabis, D.: A general reinforcement learning algorithm that masters
  chess, shogi, and go through self-play.
\newblock Science \textbf{362}(6419), 1140--1144 (2018).
\newblock \doi{10.1126/science.aar6404}.
\newblock \urlprefix\url{https://doi.org/10.1126%2Fscience.aar6404}

\bibitem{[2]}
Simon, H.A.: Why should machines learn?
\newblock In: {Michalski R. S., Carbonell J. G., Mitchell T. M. (Eds.) Machine
  Learning}, pp. 25--37. Morgan Kaufmann (1983)

\bibitem{[15]}
Simpson, E.H.: The interpretation of interaction in contingency tables.
\newblock Journal of the Royal Statistical Society: Series B (Methodological)
  \textbf{13}(2), 238--241 (1951)

\bibitem{[53]}
Solomonoff, R.J.: The time scale of artificial intelligence: Reflections on
  social effects.
\newblock Human Systems Management \textbf{5}(2), 149--153 (1985)

\bibitem{[120]}
Srivastava, N., Hinton, G., Krizhevsky, A., Sutskever, I., Salakhutdinov, R.:
  Dropout: a simple way to prevent neural networks from overfitting.
\newblock The journal of machine learning research \textbf{15}(1), 1929--1958
  (2014)

\bibitem{stekhoven2012}
Stekhoven, D.J., B{\"u}hlmann, P.: Missforest—non-parametric missing value
  imputation for mixed-type data.
\newblock Bioinformatics \textbf{28}(1), 112--118 (2012)

\bibitem{[56]}
Sutton, R.S., Barto, A.G.: Reinforcement learning: An introduction.
\newblock MIT press (2018)

\bibitem{[22]}
Teichmann, M., Weber, M., Zoellner, M., Cipolla, R., Urtasun, R.: Multinet:
  Real-time joint semantic reasoning for autonomous driving.
\newblock In: 2018 IEEE Intelligent Vehicles Symposium (IV), pp. 1013--1020.
  IEEE (2018)

\bibitem{[140]}
{The Economist}:
  \url{https://www.economist.com/leaders/2017/05/06/the-worlds-most-valuable-resource-is-no-longer-oil-but-data}
  (2017).
\newblock Accessed 27.04.2020

\bibitem{[122]}
Theodorou, V., Abell{\'{o}}, A., Thiele, M., Lehner, W.: Frequent patterns in
  {ETL} workflows: An empirical approach.
\newblock Data {\&} Knowledge Engineering \textbf{112}, 1--16 (2017).
\newblock \doi{10.1016/j.datak.2017.08.004}.
\newblock \urlprefix\url{https://doi.org/10.1016%2Fj.datak.2017.08.004}

\bibitem{[114]}
Tibshirani, R.: Regression shrinkage and selection via the lasso.
\newblock Journal of the Royal Statistical Society Series B Statistical
  Methodology \textbf{58}(1), 267--288 (1996).
\newblock \doi{10.1111/j.2517-6161.1996.tb02080.x}.
\newblock \urlprefix\url{https://doi.org/10.1111%2Fj.2517-6161.1996.tb02080.x}

\bibitem{[117]}
Tibshirani, R.: {The {LASSO} Method for Variable Selection in the Cox Model}.
\newblock Statistics in Medicine \textbf{16}(4), 385--395 (1997).
\newblock
  \doi{10.1002/(sici)1097-0258(19970228)16:4<385::aid-sim380>3.0.co;2-3}.
\newblock
  \urlprefix\url{https://doi.org/10.1002%2F%28sici%291097-0258%2819970228%2916%3A4%3C385%3A%3Aaid-sim380%3E3.0.co%3B2-3}

\bibitem{[138]}
Valiant, L.G.: A theory of the learnable.
\newblock Communications of the {ACM} \textbf{27}(11), 1134--1142 (1984).
\newblock \doi{10.1145/1968.1972}.
\newblock \urlprefix\url{https://doi.org/10.1145%2F1968.1972}

\bibitem{[139]}
Valiant, L.G.: Probably approximately correct: nature's algorithms for learning
  and prospering in a complex world.
\newblock Choice Reviews Online \textbf{51}(05), 51--2716--51--2716 (2013).
\newblock \doi{10.5860/choice.51-2716}.
\newblock \urlprefix\url{https://doi.org/10.5860%2Fchoice.51-2716}

\bibitem{vanbuuren2018}
Van~Buuren, S.: Flexible imputation of missing data.
\newblock CRC Press (2018)

\bibitem{[57]}
Vapnik, V.: Statistical learning theory.
\newblock Wiley New York (1998)

\bibitem{[48]}
Wager, S., Athey, S.: Estimation and inference of heterogeneous treatment
  effects using random forests.
\newblock Journal of the American Statistical Association \textbf{113}(523),
  1228--1242 (2018)

\bibitem{[110]}
Wager, S., Hastie, T., Efron, B.: Confidence intervals for random forests: The
  jackknife and the infinitesimal jackknife.
\newblock The Journal of Machine Learning Research \textbf{15}(1), 1625--1651
  (2014)

\bibitem{[115]}
Wager, S., Wang, S., Liang, P.S.: Dropout training as adaptive regularization.
\newblock In: Advances in neural information processing systems, pp. 351--359
  (2013)

\bibitem{warner1996understanding}
Warner, B., Misra, M.: Understanding neural networks as statistical tools.
\newblock The American Statistician \textbf{50}(4), 284--293 (1996)

\bibitem{weihs}
Weihs, C., Ickstadt, K.: Data science: the impact of statistics.
\newblock International Journal of Data Science and Analytics \textbf{6}(3),
  189--194 (2018)

\bibitem{[124]}
Wikipedia:
  \url{https://en.wikipedia.org/wiki/Simpson%27s_paradox#/media/File:Simpson's_paradox_continuous.svg}
  (2020).
\newblock Accessed 28.07.2020

\bibitem{[6]}
Wired.com:
  \url{https://www.wired.com/story/ubers-self-driving-car-didnt-know-pedestrians-could-jaywalk/}.
  (2019).
\newblock Accessed 13.05.2020

\bibitem{[13]}
Wolf, M.J., Miller, K., Grodzinsky, F.S.: Why we should have seen that coming:
  comments on microsoft's tay experiment, and wider implications.
\newblock ACM SIGCAS Computers and Society \textbf{47}(3), 54--64 (2017)

\bibitem{[75]}
Zacharaki, E.I., Wang, S., Chawla, S., Yoo, D.S., Wolf, R., Melhem, E.R.,
  Davatzikos, C.: Classification of brain tumor type and grade using {MRI}
  texture and shape in a machine learning scheme.
\newblock Magnetic Resonance in Medicine \textbf{62}(6), 1609--1618 (2009).
\newblock \doi{10.1002/mrm.22147}.
\newblock \urlprefix\url{https://doi.org/10.1002%2Fmrm.22147}

\bibitem{[116]}
Zaremba, W., Sutskever, I., Vinyals, O.: Recurrent neural network
  regularization.
\newblock arXiv preprint arXiv:1409.2329v5  (2014)

\bibitem{[136]}
Zhu, J., Chen, J., Hu, W., Zhang, B.: Big learning with bayesian methods.
\newblock National Science Review \textbf{4}(4), 627--651 (2017).
\newblock \doi{10.1093/nsr/nwx044}.
\newblock \urlprefix\url{https://doi.org/10.1093%2Fnsr%2Fnwx044}

\end{thebibliography}

\end{document}